\documentclass[%
superscriptaddress,
reprint,
amsmath,amssymb,
aps,longbibliography
]{revtex4-1}

\usepackage{graphicx}
\usepackage{dcolumn}
\usepackage{bm}
\usepackage[mathlines]{lineno}
\usepackage[bottom]{footmisc}
\usepackage{booktabs}
\makeatletter
\renewcommand*{\@fnsymbol}[1]{\ensuremath{\ifcase#1\or *\or ** \or \ddagger\or
		\mathsection\or \mathparagraph\or \|\or **\or \dagger\dagger
		\or \ddagger\ddagger \else\@ctrerr\fi}}
\makeatother

\begin{document}

	\title{Inverse design of ultracompact multi-focal optical devices by diffractive neural networks}
	
	\author{Yuyao Chen}
	\affiliation{Department of Electrical and Computer Engineering and Photonics Center, Boston University, 8 Saint Mary’s Street, Boston, MA, 02215, USA}
	
	\author{Yilin Zhu}
	\affiliation{Division of Materials Science and Engineering, Boston University, 15 St Mary’s St, Brookline, MA, 02246, USA}
	
	\author{Wesley A. Britton}
	\affiliation{Division of Materials Science and Engineering, Boston University, 15 St Mary’s St, Brookline, MA, 02246, USA}

	\author{Luca Dal Negro}
	\email[email:]{dalnegro@bu.edu}
	\affiliation{Department of Electrical and Computer Engineering and Photonics Center, Boston University, 8 Saint Mary’s Street, Boston, MA, 02215, USA}
	\affiliation{Division of Materials Science and Engineering, Boston University, 15 St Mary’s St, Brookline, MA, 02246, USA}
	\affiliation{Department of Physics, Boston University, 590 Commonwealth Avenue, Boston, MA, 02215, USA}
	
	\begin{abstract}
		We propose an efficient inverse design approach for multifunctional optical elements based on adaptive deep diffractive neural networks (a-D$^2$NNs). Specifically, we introduce a-D$^2$NNs and design two-layer diffractive devices that can selectively focus incident radiation over two well-separated spectral bands at desired distances. We investigate focusing efficiencies at two wavelengths and achieve targeted spectral lineshapes and spatial point-spread functions (PSFs) with optimal focusing efficiency. In particular, we demonstrate control of the spectral bandwidths at separate focal positions beyond the theoretical limit of single-lens devices with the same aperture size. Finally, we demonstrate devices that produce super-oscillatory focal spots at desired wavelengths. The proposed method is compatible with current diffractive optics and doublet metasurface technology for ultracompact multispectral imaging and lensless microscopy applications.
	\end{abstract}
	
	\maketitle
	
	Multifunctional diffractive optical elements (DOEs), when integrated atop on-chip detectors, enable ultracompact imaging functionalities for miniaturized flat cameras and microscopes \cite{Banerji:19,britton2020phase,boominathan2020phlatcam,Banerji:20}. Multispectral behavior is often achieved by partitioning single layer devices into separate phase regions that affect different wavelengths. However, this design limits the maximum efficiency achievable at each wavelength, which is a significant challenge for DOEs working at multiple wavelengths \cite{lin2016photonic,arbabi2016multiwavelength}. This is because when one specific wavelength illuminates the entire device, only the phase region designed to operate at that wavelength will produce the desired output while the other part of the illuminated device area will not, thus requiring a different approach.
	
	In order to address this important challenge, we propose here novel multi-layer designs based on the flexibility of adaptive diffractive neural networks (a-D$^2$NNs) for the engineering of multi-layered diffractive devices with targeted spectral response and spatial point-spread functions (PSFs) at different wavelengths. Recently, deep diffractive neural networks (D$^2$NNs) that combine optical diffraction with deep learning capabilities have been reported and applied to all-optical diffraction-based systems that implement object recognition \cite{lin2018all}. Moreover, D$^2$NNs have also been demonstrated successfully for the inverse design of multi-layered diffractive devices that achieve pulse shaping \cite{veli2021terahertz} and broadband filtering \cite{luo2019design}. These devices are macroscopic with typical dimensions up to the $cm$ size and are fabricated using 3D printing for applications in the Terahertz domain \cite{luo2019design,veli2021terahertz}. However, the design of diffractive devices that over multiple spectral bands in the optical regime is very challenging and requires a more flexible implementation of the D$^2$NNs platform. 
	
	In this paper, we introduce and utilize a-D$^2$NNs that leverage an adaptive loss weight algorithm for the inverse design of two-layer, ultracompact dual-band DOEs. The a-D$^2$NNs are trained to maximize the focusing efficiencies for $\lambda_1$ at $f_1$ and $\lambda_2$ at $f_2$. The engineered devices show efficiencies over $50\%$ at both targeted wavelengths, which exceeds the limit of phase-modulated single layer DOEs  \cite{lin2016photonic,arbabi2016multiwavelength,britton2020compact}. We systematically investigate how the focusing efficiencies vary with the distance between the two diffractive layers and the pixel size, taking into account practical fabrication constraints. We also investigate how the efficiency is affected by the phase discretization level of the proposed diffractive devices. In addition, the obtained phase designs can also be implemented using current metasurface technology \cite{Khorasaninejadeaam8100,IPRNA,Banerji:19,Yilmaz_2019}, including the recently developed doublet metasurface fabrication approach \cite{groever2017meta,Augusto2022Fundamental}. An important aspect of our approach is the design of the spectral lineshapes of DOEs. In fact, we demonstrate dual-band devices with designed bandwidths that are narrower compared to diffractive lenses with the same aperture size. Finally, we show that a-D$^2$NNs can be implemented to design devices with desired spatial PSFs, including DOEs that produce super-oscillatory fields with focal spots below the diffraction limit \cite{rogers2013optical}.
	
	\begin{figure}[htbp!]
		\centering\includegraphics[width=\linewidth]{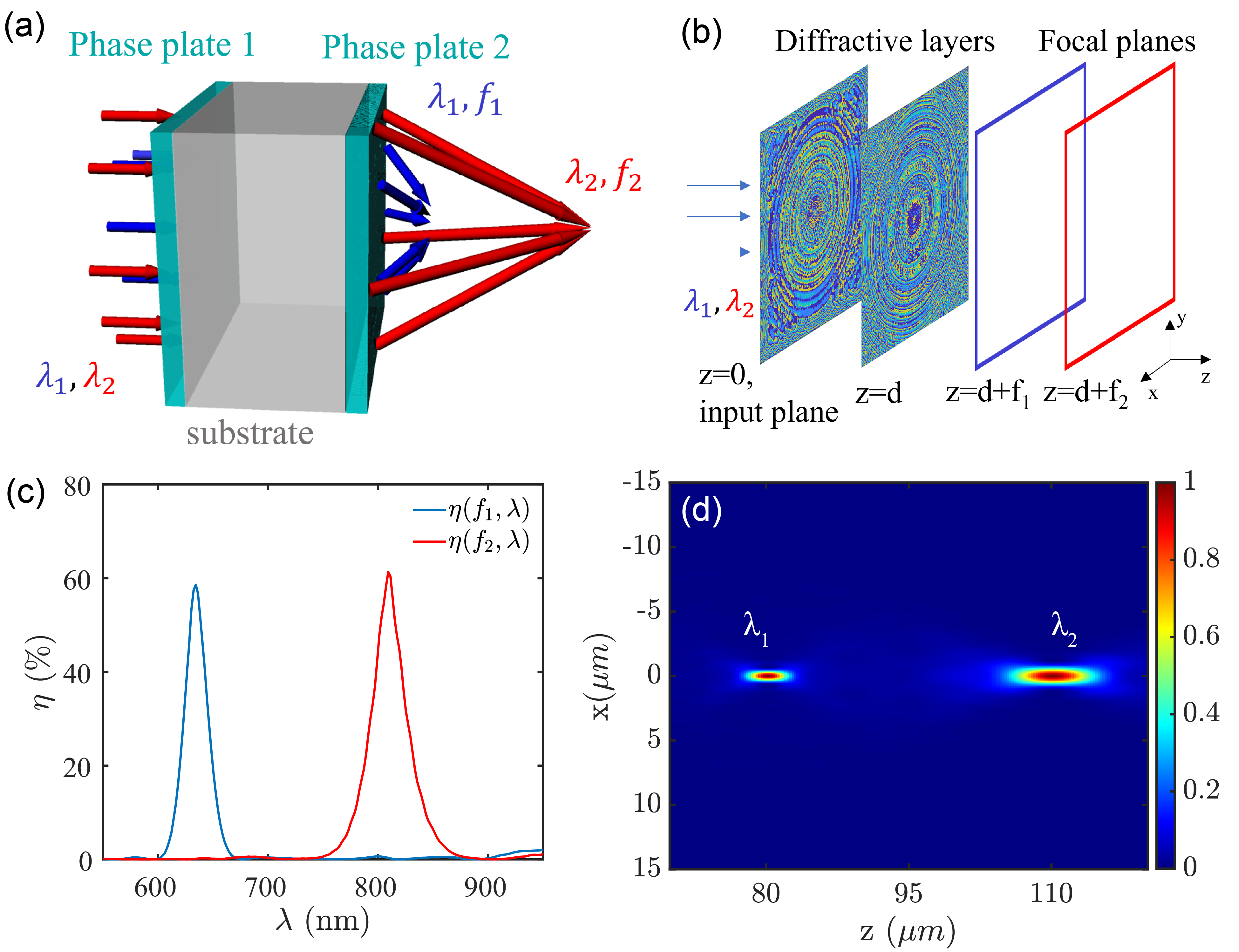}
		\caption{(a) Schematics of a two-layer dual-band DOE and (b) D$^2$NN representation. (c) Focusing efficiency spectra for the device in panel (a) with a-D$^2$NN. (d) Side view of normalized diffraction intensity at $\lambda_1$ and $\lambda_2$.}
		\label{Fig1}
	\end{figure}
	
	Figure \ref{Fig1} (a) illustrates the general two-layer diffractive device concept consisting of two diffractive phase plates located on both sides of a transparent substrate. The varying thickness profiles of the materials on the phase plates impart different phase shifts to the waves that propagate through the device. A schematic design of the a-D$^2$NN that implements such a device is shown in Fig. \ref{Fig1} (b), where the two diffractive layers of the a-D$^2$NN correspond to the two phase plates of the device. We implement the Rayleigh-Sommerfeld (RS) first integral formulation within the a-D$^2$NN in order to simulate the forward light propagation from one plane to the next one, according to the model \cite{britton2020compact,britton2020phase}:
	\begin{eqnarray}\label{RS equation}
	A_{o}\left(x^\prime,y^\prime\right)= A_{s}\left(x,y\right) * h(x,y;x^\prime,y^\prime;z,k)\\
	h(x,y;z,k)=\frac{1}{2\pi}\frac{d}{r}  \left(\frac{1}{r}-jk\right)\frac{e^{jkr}}{r}.
	\end{eqnarray}
	where $*$ denotes the two-dimensional spatial convolution, $A_{o}$, $A_{s}$ are the transverse field distributions on the source and observation plane with coordinates $(x,y)$ and $(x^\prime,y^\prime)$, respectively. Moreover, $k=\frac{2\pi{n}}{\lambda}$ is the wave number, where $\lambda$ is the incident wavelength in vacuum and $n$ is the index of medium between the two planes. We use $r=\sqrt{x^2+y^2+z^2}$, where $z$ is the distance between the two planes. In our two-layer DOE, we first compute the forward propagation from plane $z=0$ to plane $z=d$ at wavelengths $\lambda_1$ and $\lambda_2$. Then the field distributions on the focal plane for $\lambda_1$ at $z=d+f_1$ and for $\lambda_2$ at $z=d+f_2$ are calculated. We then utilize the a-D$^2$NN to maximize the focusing efficiency $\eta$ at these two focal planes, using the following definition for the focusing efficiency \cite{britton2020compact}:
	\begin{equation}
	\eta(\lambda, z)=\frac{\int_{0}^{2 \pi} \int_{0}^{3 \mathrm{FWHM}/2} I^{\prime}\left(\lambda, z, \rho^{\prime}, \theta^{\prime}\right) d r^{\prime}  d \theta^{\prime} }{\iint I(\lambda, z=0, \rho, \theta) dS}
	\end{equation}
	where $I^{\prime}$ denotes the intensity distribution on the focal plane, $I$ denotes the one on the input plane, and $S$ denotes the input plane aperture. The symbols $(\rho,\theta)$ and $(\rho',\theta')$ are the polar coordinates on the focal and input plane, respectively.

	\begin{figure}[t!]
		\centering\includegraphics[width=\linewidth]{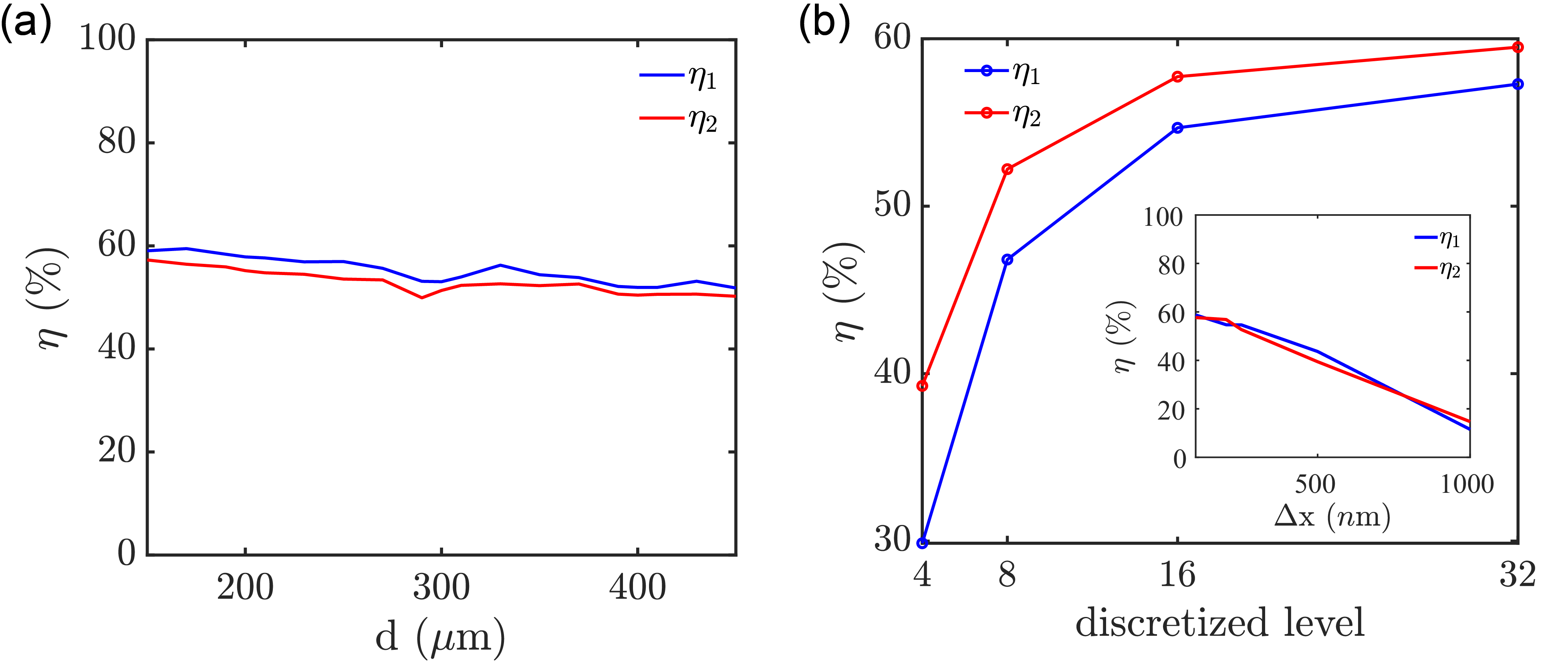}
		\caption{(a) Focusing efficiency with respect to the distance $d$ between the two diffractive layers (b) Focusing efficiency with respect to the number of discrete phase levels. The inset shows the dependence of the focusing efficiencies on the pixel size (minimum spatial feature of the phase profile).}
		\label{Fig2}
	\end{figure}
	
	The focusing efficiency is utilized in the loss function of the a-D$^2$NN as follows:
	\begin{equation}\label{focusing efficiency loss funciton}
	\mathcal{L} = w_1(1-\eta_1)^2+ w_2(1-\eta_2)^2
	\end{equation}
	where $\eta_1=\eta(\lambda_1,d+f_1)$, $\eta_2=\eta(\lambda_2,d+f_2)$, and $w_1$ and $w_2$ are the loss weights. Based on the definition of a suitable loss function, the a-D$^2$NN is directly trained using error backpropagation within the diffractive layers without the need of training datasets. Therefore, the a-D$^2$NN achieves a more efficient inverse design of complex phase devices compared to data-driven neural network approaches \cite{ma2021deep,liu2021tackling}. Specifically, a-D$^2$NN are trained by varying the phase profiles on the two diffractive layers in order to minimize $\mathcal{L}$. We train the latent variable $h_\ell$ on each pixel of the diffractive layers related to the material thickness $h$ by $h=h_{max}(\sin(h_\ell)+1)/2$, where $h_{max}$ is the specified maximum thickness of the device. The phase profile $\phi(x,y)$ induced by diffractive layers at wavelength $\lambda$ is $\phi=\frac{2\pi}{\lambda}(n-1)h$. As a proof of concept, we select $\lambda_1=632.8~nm$, $\lambda_2=808~nm$, $f_1=80~{\mu}m$, and $f_2=110~{\mu}m$, which were used in our previous work \cite{britton2020compact}. The two diffractive layers are square apertures with $L=100~{\mu}m$ side length and the device pixel size is $\Delta x=200~nm$. We assume that the substrate index $n=3$, $h_{max}=500~nm$, and thickness $d=200~\mu m$. Differently from the usual D$^2$NN approach, here we implement adaptive loss weights that balance the interplay between different loss terms automatically depending on their values \cite{wang2020understanding,chen2021Physics}. In particular, we apply the following updates for $w_m~(m=1,2)$ at the $k^{th}$ epoch of training: 
	\begin{equation}
	w_m^{k} \leftarrow w_m^{k-1} + \gamma(1-\eta_m)^2     
	\end{equation}
	where $\gamma$ is the learning rate for loss weights update and we choose $\gamma=1$. We use a desktop with GeForce GTX 1080 Ti graphical processing unit (GPU, Nvidia Inc.), an Intel i7-8700K central processing unit (CPU, Intel Inc.) and 32 GB of RAM for training. We trained the a-D$^2$NN over $2000$ epochs using Adam optimizer with learning rate equals to $0.1$. The typical training time is only $\sim 10$ minutes. In Fig. \ref{Fig1} (c) we show the obtained focusing efficiency spectra of the device at $f_1$ and $f_2$. Specifically, we observe that $\eta(f_1,\lambda)$ and $\eta(f_2,\lambda)$ are peaked at $\lambda_1$ and $\lambda_2$, respectively, and that both $\eta_1$ and $\eta_2$ values exceed $50\%$. Therefore, the designed two-layer dual-band device exceeds the efficiency limit expected in a single-layer DOE. Moreover, in Fig. \ref{Fig1} (d) we display the side view of the normalized intensity diffraction of the device, clearly showing that the two targeted wavelengths $\lambda_1$ and $\lambda_2$ are well focused at the designed focal lengths $f_1$ and $f_2$, respectively. 
	
	We further investigate the influence of the distance $d$ between the two diffractive layers while keeping all the other parameters as specified above. The obtained $\eta_1$ and $\eta_2$ for devices with different $d$ are shown in \ref{Fig2} (a), which demonstrates that for all the devices the values of $\eta_1$ and $\eta_2$ remain above $50\%$. We next consider the discretization of the obtained continuous phase profiles into discrete levels used for scalable lithographic fabrication \cite{britton2020compact,britton2020phase}. Figure. \ref{Fig2} (b) displays how the number of discretized levels of the device affects $\eta_1$ and $\eta_2$. In particular, we find that $\eta_1$ and $\eta_2$ increase with the number of phase levels used, with a similar scaling to the one reported for the diffraction efficiency of multi-level gratings \cite{swanson1989binary}. In particular, we observe that our two-layer device can achieve $\eta_1=46\%$ and $\eta_2=52\%$ when only 8 discretization levels are used. The inset of Fig. \ref{Fig2} (b) displays the focusing efficiency with respect to the pixel size $\Delta x$. Devices with smaller $\Delta x$ achieve larger $\eta_1$ and $\eta_2$ efficiencies as they accommodate faster phase variations. Therefore, our analysis indicates that the proposed devices can be conveniently engineered using current multi-level DOE technology \cite{britton2020compact,britton2020phase} as well as planar metasurfaces that provide nanoscale phase resolution ($50~nm\sim 300~nm$) \cite{Khorasaninejadeaam8100,IPRNA,Banerji:19,groever2017meta}.

	\begin{figure}[t!]
		\centering\includegraphics[width=\linewidth]{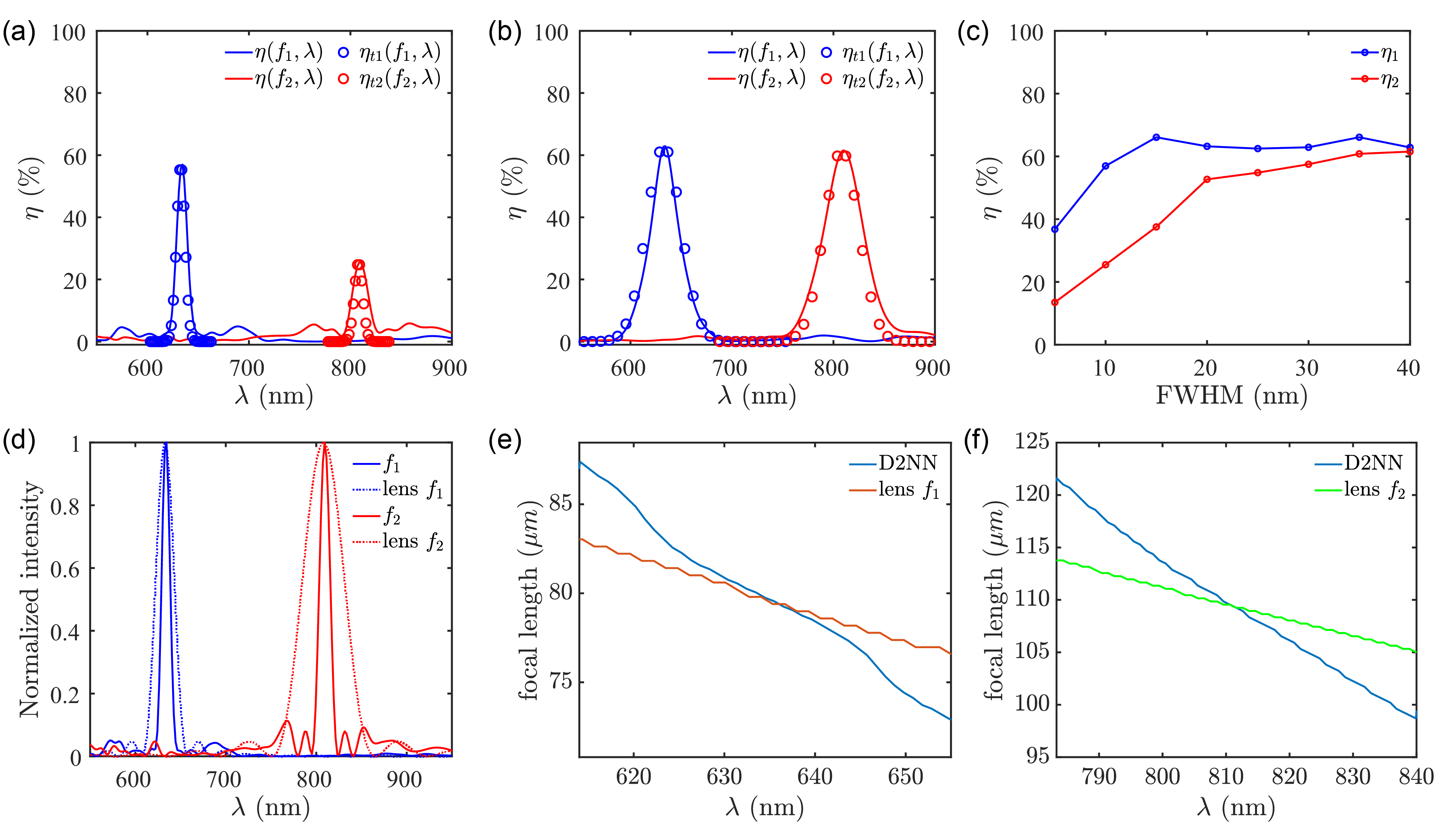}
		\caption{Spectral lineshapes with (a) $\sigma=10~nm$ and (b) $\sigma=40~nm$. (c) $\eta_1$ and $\eta_2$ with respect to $\sigma$. (d) Normalized field intensity spectra for $\sigma=10~nm$ at $f_1$ (solid blue) and $f_2$ (solid red) compared to the single diffractive lenses with focus $\lambda_1$ at $f_1$ (dashed blue) and $\lambda_2$ at $f_2$ (dashed red). (e-f) Wavelength dependence of the focal lengths at $\lambda_1$ and $\lambda_2$, respectively.}
		\label{Fig3}
	\end{figure}
	
	Another important advantage of the DOE design based on a-D$^2$NNs is that we can engineer spectral lineshapes by modifying the loss function used for training the network. In order to demonstrate this capability, we train a-D$^2$NNs to obtain devices with spectral lineshapes for the focusing efficiency $\eta_{tm}(f_m,\lambda)\quad (m=1,2)$ described by the expression:
	\begin{equation}
	\eta_{tm}(f_m,\lambda) = \exp\left[-4\log(2)\left(\frac{\lambda-\lambda_m}{\sigma_m}\right)^2\right]\quad (m=1,2)   
	\end{equation}
	where $\lambda_m$ and $\sigma_m$ ($m=1,2$) are the center wavelength and FWHM of the targeted Gaussian spectral lineshape, respectively. We modify the loss function in a-D$^2$NN as follows:
	\begin{equation}
	\begin{split}
	\mathcal{L} =& \sum_{m=1,2} \sum^N_{k=1} w_m(1-\eta_m)^2+ \frac{1}{N}w_{sm}\left[\eta_m* \eta_{tm}(f_m,\lambda^m_k)\right. \\& -\left.\eta_{m}(f_m,\lambda^m_k)\right]^2     
	\end{split}
	\end{equation}
	where $w_{sm}~(m=1,2)$ is the loss weight for the spectral lineshape loss term. The first term is the same used in Eq. \ref{focusing efficiency loss funciton}. For the second term, we sample $N$ discrete wavelengths of the target spectrum uniformly from $\lambda^m_{min}$ to $\lambda^m_{max}$ centered at $\lambda_m~(m=1,2)$ and evaluate the focusing efficiencies over these wavelengths. The mean squared error (MSE) between the obtained $\eta_{m}(f_m,\lambda^m_k)$ and the target lineshape $\eta_{tm}(f_m,\lambda)$ with its maximum rescaled to $\eta_m$ is then calculated. During the training process, we apply the adaptive loss weights for both $w_m$ and $w_{sm}$. We train the a-D$^2$NN with the same parameters used to generate Fig. \ref{Fig1} and we fix $\sigma_1=\sigma_2=\sigma$. In particular, we sampled $N=30$ wavelengths over two ranges between $\lambda^m_{min}=\lambda_m-3\sigma$ and $\lambda^m_{max}=\lambda_m+3\sigma$. The a-D$^2$NN is trained over $2000$ epochs. We show the spectral results for the device trained using $\sigma=10~nm$ and $\sigma=40~nm$ in Fig. \ref{Fig3} (a) and (b), respectively. Furthermore, in Fig. \ref{Fig3} (c) we display how $\eta_1$ and $\eta_2$ vary for devices optimized with different $\sigma$. A sharp drop of focusing efficiency is observed when the width of the targeted Gaussian lineshape is decreasing below $20~nm$. We also evaluate the normalized field intensity spectra at the origin of focal planes at $f_1$ and $f_2$ for the device with $\sigma=10~nm$. We compare our results with the ones of two diffractive lenses with the same dimension that focus $\lambda_1$ at $f_1$ and $\lambda_2$ at $f_2$. The analytical expression for the normalized intensity spectrum $I_m(\lambda)$ of a diffractive lens that focuses $\lambda_m$ at $f_m$ (m=1,2) is given by \cite{gu2000advanced}:
	\begin{equation} \label{intensity spectra}
	I_m(\lambda)=\left[\frac{\sin(u_m(\lambda)/4)}{u_m(\lambda)/4}\right]^2
	\end{equation} 
	where we defined $u_m=\frac{2\pi}{\lambda} \left(\frac{L}{2}\right)^2\left(\frac{\lambda}{f_m \lambda_m}-\frac{1}{f_m}\right)$. As shown in Fig. \ref{Fig3} (d), two-layer devices designed using the a-D$^2$NN method can feature significantly narrow bandwidths than diffractive lenses at both $\lambda_1$ and $\lambda_2$. Recalling that the intensity spectrum for diffractive lens is completely determined once the parameters $L$, $\lambda_m$, and $f_m$ are fixed, we appreciate that the designed two-layer DOEs provide the additional capability to tailor spectral lineshapes for a given aperture size. To better understand how the obtained devices achieve narrow bandwidths, we evaluate the spectral dependence of their focal lengths near $\lambda_1$ and $\lambda_2$ and compare with diffractive lenses in Fig. \ref{Fig3} (e) and (f), respectively. Our findings show that the focal lengths of the designed dual-band devices vary faster with respect to the wavelength compared to diffractive lenses. Therefore, they can achieve enhanced spectral selectivity (narrower bandwidths) due to their stronger defocusing behavior when varying the incident wavelengths. 
	
	\begin{figure}[t!]
		\centering\includegraphics[width=\linewidth]{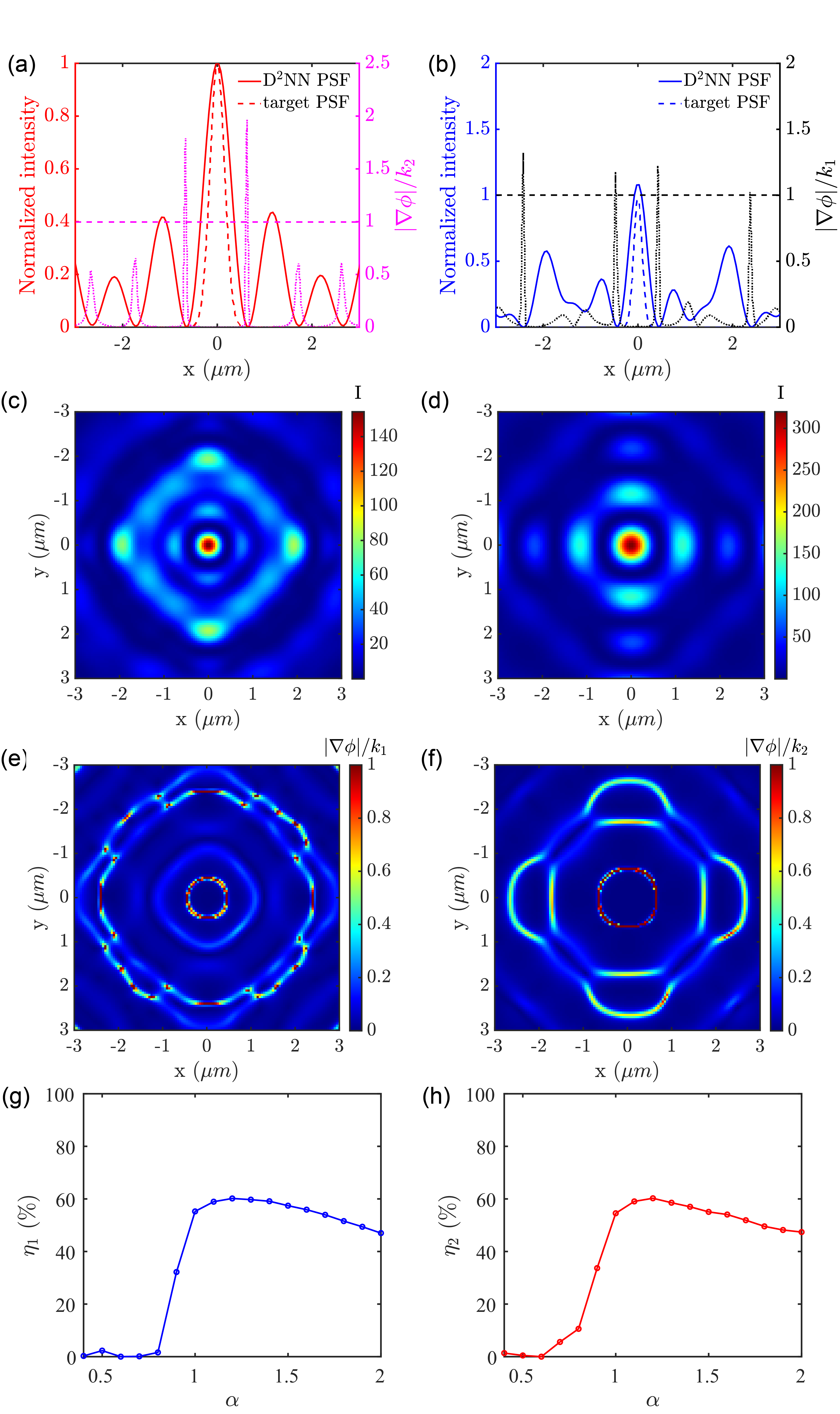}
		\caption{(a) Normalized transverse intensities at $\lambda_1$ (red solid line) and (b) at $\lambda_1$ (blue solid line) and corresponding phase gradients across the focal spots. (c, d) Focal plane intensity profiles at $\lambda_1$ and $\lambda_2$, respectively, with $\alpha=0.4$. (e, f) Normalized phase gradient maps at the focal planes for $\lambda_1$ and $\lambda_2$, respectively. For better visualization, the range of the color bars is limited to [0,1]. (g, h) $\eta_1$ and $\eta_2$ with respect to $\alpha$.}
		\label{Fig4}
	\end{figure}
	
	We finally implement a-D$^2$NNs for the inverse design of two-layer DOEs with desired focusing PSF. We model the PSF by two-dimensional Gaussian function on the focal plane $f_m~(m=1,2)$:
	\begin{equation} \label{spatial intensity}
	I_{tm}(x,y,z_m,\lambda_m)= \exp\left[-4\log(2)\left(\frac{x^2+y^2}{(\alpha \epsilon_m) ^2}\right)\right]
	\end{equation} 
	where $z_m=d+f_m$ are the focal plane $z-$ coordinate for $\lambda_m$, $\alpha$ is a scaling constant that quantifies the degree of spatial localization of the designed focal spot with respect to the Rayleigh diffraction limit, which is achieved for $\alpha=1$, and $\epsilon_m=0.51\frac{\lambda_m f_m }{L}$ is the diffraction limited FWHM of the focal spot. In order to obtain desired spatial PSFs we implement the following loss function for training:
	\begin{equation}
	\begin{split}
	\mathcal{L} &= \sum_{m=1,2} \sum_{x,y} w_m(1-\eta_m)^2+ w_{pm}\left[I_m(0,0,z_m,\lambda_m)\right.\\
	&\left.*I_{tm}(x,y,z_m,\lambda_m)-I_m(x,y,z_m,\lambda_m)\right]^2     
	\end{split}
	\end{equation}
	where $w_{pm}$ is the loss weight for the loss term of squared error between device real PSF $I_m(x,y,z_m,\lambda_m)$ and targeted PSF $I_{tm}(x,y,z_m,\lambda_m)$ with its maximum rescaled to $I_m(0,0,z_m,\lambda_m)$. In particular, we trained a-D$^2$NN using $\alpha=0.4$, which corresponds to a FWHM below the diffraction limit. The intensity cuts through the center of the focal spots at $\lambda_1$ and $\lambda_2$ are shown in Fig. \ref{Fig4} (a) and (b), respectively. The dashed lines are the targeted PSF used for training the network. The obtained intensity profiles indicate the formation of optical super-oscillations, which have been shown to result in arbitrarily small energy concentration without the assistance of evanescent waves \cite{berry2006evolution,ferreira2006superoscillations}. We note that the obtained PSFs exhibit the presence of significant sidebands compared to the targeted Gaussian PSF. This is due to the fundamental nature of super-oscillations in which enhanced (sub-diffractive) field focusing can only be achieved  at the expense of a polynomial increase in the power directed into the sidebands \cite{rogers2013optical}. Due to their extreme localization properties, optical super-oscillations have found applications to sub-wavelength imaging and microscopy \cite{rogers2012super}. In order to demonstrate super-oscillations in our devices we studied the phase gradient $|\nabla \phi|$ of the diffracted field on the focal plane, which corresponds to a local wave number. Super-oscillations form when $|\nabla \phi|>k_m$, where $k_m=\frac{2\pi}{\lambda_m} (m=1,2)$ are the incident wave numbers. In Fig. \ref{Fig4} (a) and (b) we display for the two wavelengths of interest the phase gradient profiles of the fields normalized by $k_m$. We notice that the peaks of $|\nabla \phi|/k_m$ exceed unity around the designed focal spots, demonstrating the super-oscillation character of the waves. We further show the two-dimensional focal intensity distributions and $|\nabla \phi|/k_m$ maps on the two different focal planes in Fig. \ref{Fig4} (c-d) and (e-f), respectively. In Fig. \ref{Fig4}(g) and (h) we summarize our results for the variation of focusing efficiencies with respect to the localization parameter $\alpha$. We note that the focusing efficiencies slightly decrease when increasing $\alpha$ if $\alpha>1$ while they suddenly drop to almost zero by decreasing $\alpha$ when $\alpha<1$, consistently with the super-oscillating regime \cite{berry2006evolution,ferreira2006superoscillations}. 
	
	In conclusion, we introduced an inverse design approach for dual-band multi-focal DOE based on flexible a-D$^2$NNs. We demonstrate novel two-layer designs that show $\eta_1,\eta_2>50\%$, beyond the limit of single-layer DOEs working at two wavelengths. Furthermore, we showed the designs of DOEs with desired spectral lineshapes and FWHM down to $\sigma=5~nm$. Finally, we show PSF engineering with designed super-oscillatory focal spots below the diffraction limit. The flexible approach introduced here enables the engineering of two-layer diffractive devices with desired spectral and spatial responses for multi-band imaging and microscopy applications.
	
	\section*{Funding Information}
	National Science Foundation (ECCS-2015700).

	\section*{Disclosures}
	The authors declare no conflicts of interest.

\end{document}